\documentclass{ws-procs961x669}            % default, citations in superscript
\begin{document}
\title{
Massive Primordial Black Holes
in Contemporary and Young Universe\\
(old predictions and new data)}
\author{A.D. Dolgov$^*$}

\address
{Physics Department, Novosibirsk State University,\\
Novosibirsk, 630090, Russia\\
$^*$E-mail: dolgov@fe.infn.it
}

\begin{abstract}

A brief review of the recent astronomical data, indicating that the universe is abundantly populated by
heavy black holes (BH), is presented.  Conventional astrophysics and cosmology cannot explain such a high
population of BHs. A mechanism of the paper of 1963 is described, which at least qualitatively explained  the  
observational data. In particular, the prediction that massive primordial BHs can  be cosmological dark matter
"particles"  is discussed. 

\end{abstract}


\bodymatter

\section{Introduction}\label{s-intro}
{In this talk two different but related subjects are considered:}\\
{\it I. A mechanism of massive PBH formation, which is  much different from the caninocal ones.}
{According to this mechanism  the fundament of PBH creation is build at inflation by making large isocurvature
fluctuations at relatively small scales, with practically vanishing density perturbations.}  
{It is achieved by a simple modification of a popular scenario of baryogenesis.
Density perturbations are generated rather late after the QCD phase transition.}
{The emerging universe looks like a piece of Swiss cheese, where (the black)
holes are high baryonic 
density objects occupying a minor fraction of the universe volume.}\\
{{\it II. A brief review of new (and not only new) astronomical data which are in strong tension with the accepted 
standard  cosmological model.}} 
{The data nicely fit the suggested scenario of PBH formation.
More and more observational evidence in favor of massive and supermassive PBH 
{and other, not yet explained  in the standard way phenomena}  in the universe, appears practically every week.

{The usual or astrophysical  black holes are the results of stellar evolution after a star exhausted  its nuclear fuel
and collapsed into a compact object, either into a neutron star or into a black hole. This collapse into black hole happens 
if the mass of the star 
is, roughly speaking, larger than three 
solar masses. Such black holes can be created in sufficiently old or even contemporary universe.}
%{Masses are of the order of a few solar masses, }\\

{There exist also the astrophysical supermassive black holes (SMBH) with huge masses, ${M \sim(10^6 - 10^{9}) M_\odot}$,
where $M_\odot \approx 2 \times 10^{33}  $ g is the solar mass.
They are supposed to be the products of matter accretion
to smaller BHs or matter accretion to matter excess in galactic centers.}

Primordial black holes (PBH) formed in the very early universe if the density excess at the cosmological 
horizon was large, ${\delta \rho /\rho \gtrsim 1}$, as it was  first suggested by Zeldovich and 
Novikov~\cite{ZN}. {Normally the masses of PBH created through this mechanism were supposed to be 
rather low and the mass spectrum was quite sharp, close to the delta-function.}

An alternative mechanism of formation of very massive PBH  with  wide spread mass spectrum was proposed in 
ref{. \cite{ADJS}, see also the subsequent paper~\cite{DKK} , where the model was further elaborated.
{Heretic declaration of 1993}  is turning now into the acknowledged faith with  more and more
astronomical data proving its truth.
The conclusion that the observed mysterious phenomena are induced by {\it  massive primordial} black holes
allows to cure an avalanche of inconsistencies with the standard ${\Lambda}$CDM cosmology and astrophysics.

In the most extreme form the suggested in 1993 scenario gives rise to:\\
{$\bullet$ {Cosmological Dark Matter fully (or predominantly) made by PBHs.\\
%Primordial BHs make all or dominant part of dark matter (DM).}}\\
{$\bullet$ {Primordial creation of majority of black holes in the universe.}}\\
{$\bullet$ {Very early QSO formation}.}\\
{$\bullet$ {Early creation of metals and dust}.}\\
{$\bullet$ {Seeding of large galaxies by supermassive PBH.}}\\
$\bullet$ {Seeding of globular clusters by ${10^3 - 10^4}$ PBHs} 
and dwarf galaxies by ${10^4 - 10^5}$ PBHs~\cite{AD-KP}.\\
{$\bullet$ {Clouds of matter with high baryon-to-photon ratio.}}\\
$\bullet$ {A possible by-product: plenty of (compact) anti-stars, even in the Galaxy.}

Due to lack of space many references are omitted here. They can be found in the reviews~\cite{monsters,AD-UFN}.
together with more detailed discussion %can be found in the reviews~\cite{monsters,AD-UFN}.

\section{Brief review of old and recent observations and discoveries \label{s-rev-obs}}

\subsection{Data from $z\sim 10$ universe \label{ss-z-hi}}

{The observations of the last several years indicate that  the young universe at ${z \sim 10}$ is grossly overpopulated 
with unexpectedly high amount of:} \\
{1. Bright QSOs, alias supermassive BHs, up to ${M \sim 10^{10} M_\odot}$,}\\
{2. Superluminous young galaxies,}\\
{3. Supernovae, gamma-bursters,}\\
4. Dust and heavy elements.\\
These facts are in good agreement with the predictions listed in the previous page,} 
but in conflict with the expectations of the Standard Cosmological Model (SCM) at least  by an order of magnitude.

%Some examples of the data.\\ 
{\it 1. Supermassive BH, or QSO.}\\
About 40 quasars with ${z> 6}$ are known, with BH of %\\
${ 10^9 M_\odot}$ and ${L \sim 10^{13-14} L_\odot}$,
where $L_\odot \approx 4\times 10^{33} $ erg/sec is the solar luminosity.
%\tcmag{The quasars are supposed to be supermassive black holes}
%%%\tccyan{The formation SMBH in such short time by conventional mechanisms looks problematic.}\\
{Such black holes,
{when the Universe was less than one billion years old,} 
present substantial challenges to theories of the formation and growth of
black holes and the coevolution of black holes and galaxies.}
%\tcmag{Even the origin of SMBH in contemporary universe during 14 Gyr is difficult to explain.} \\
Non-standard accretion physics and the formation of massive seeds seem to be necessary.
Neither of them is observed in the present day universe.

Two years ago another huge QSO was discovered~\cite{QSO-10}.
{There is already a serious problem with formation of lighter and less luminous quasars}
{which is multifold deepened with this new found "baby".}
{The new one with ${M \approx 10^{10} M_\odot }$  makes the formation}
{absolutely impossible in the standard approach.}
 
In ref.~\cite{accr-rate} the rate of accretion to the central black hole with $M = 10^5 M_\odot$ in the galaxy with
the mass $3\times 10^{10} M_\odot$ at the redshift $z = 7.5$ was estimated as a few times  $10^{-6} M_\odot $ per year.
So the total accreted mass was about $2000 M_\odot$ during the universe life time which at this $z$  was 320 Myr.
This is by far too little to make any of billion mass observed black holes (quasars).

Several more puzzling examples appeared  already after the conference. Among them is a black hole with the mass
{${{0.8\cdot 10^9 M_\odot}}$ in the {\it neutral} universe at ${z=7.5}$~\cite{neutral}. Neutrality of the surrounding medium means
that accretion is absent or very weak, so this BH is surely primordial.

Another recently discovered faint QSO, at $z=6$ with ${M=10^9 M_\odot}$ needs for its formation either super-Eddington accretion or 
${10^5 M_\odot}$ seed \cite{super-accr}.

{\it 2. Galaxies observed at ${z \sim 10}$:}\\
There are quite a few striking examples:
{a galaxy at {${z \approx 9.6}$}, which was created when the universe was younger than ${t_U < 0.5}$ Gyr;} 
{a galaxy at {${z \approx 11}$,} born at
%which was formed earlier than the universe age was 
$t_U \sim 0.4$ Gyr, three times more luminous in UV than other galaxies at ${z = 6-8}$.}  Recently  many others
have were discovered, which were also unexpectedly early created. Discovery of not so 
so young but extremely luminous galaxy with the luminosity
{{${L= 3\cdot 10^{14} L_\odot }$ and the age ${t_U \sim 1.3 }$ Gyr} was reported in  paper~\cite{gal-hi-L}.
{Quoting the authors:} {{"The galactic seeds, or embryonic black holes, might be bigger than thought possible.}
Or another way to grow this big is to have gone on a sustained binge, consuming food 
faster than typically thought possible". 
An essential feature of the seed is that it must have a low spin.

According to paper~\cite{gal-dens} 
the {density of galaxies at ${z \approx 11}$ is 
${10^{-6} }$ Mpc$\bm{^{-3}}$, an order of magnitude higher than estimated from the data at lower z.}
{Origin of these galaxies is unclear.}

{\it  3. Dust, supernovae, gamma-bursters...}
{Abundant dust, too much denser, then expected, 
is observed in several early  galaxies, e.g. in HFLS3 at ${ z=6.34} $ and in A1689-zD1 at ${ z = 7.55}$.} 
Catalogue of the observed dusty sources indicates that their number is an order of magnitude larger 
than predicted by the canonical theory~\cite{dust-ctlg}.

{Hence, prior to or simultaneously with the QSO formation a rapid star formation should take place.}
{These stars should evolve to a large number of
supernovae enriching interstellar space by metals through their explosions}
{which later make molecules and dust.} {(We all are dust from SN explosions, but probably at much later time.)}

{Another possibility of the evolved chemistry at high redshift is a non-standard 
big bang nucleosynthesis in the bubbles with very high baryonic density, B,
which allows for  formation of heavy elements beyond lithium.} Such high B bubbles could be created in the  frameworks of the
mechanism of refs~\cite{ADJS,DKK}.

{Observations of high redshift gamma ray bursters (GRB) also indicate 
a large abundance of supernova at redshifts close to ten.} 
{The highest redshift of the observed GBR is 9.4 and there are a few more
GBRs with smaller but still high redshifts.}\\
{The necessary star formation rate for explanation of these early
GBRs is at odds with the canonical star formation theory.}

\subsection{{ Problems of the contemporary universe: \label{ss-contempi}}} 

1. In the centers of every large galaxy a supermassive black hole (SMBH) is observed, with the mass 
close to billion solar masses in elliptic and lenticular galaxies and  a few million solar masses in
spiral galaxies, like our Milky Way.

The common faith is that these SMBHs are created by matter accretion on the density excess in the galactic
center (either a small BH, or a density cusp). However, the usual accretion is not effective enough to make
such huge BHs even in whole universe age, $t_U \sim 15 $ Gyr.

More attractive looks the inverted picture when a supermassive seed (PBH) was first formed and a galaxy is later collected
by accretion of  the usual matter surrounding it. 

It is tempting, but rather far fetched, to conclude that the type of the galaxy is determined by the properties of the seed.\\

\noindent
2. There are strikingly surprising data  indicating that SMBH are observed not only in large galaxies but also in very small
ones and even in almost {\bf empty} space,  e.g. a black hole with ${M\sim 10^9 M_\odot}$~\cite{empty-BH}.  
Seemingly these poor guys happened to be in the lower density regions of the usual matter and 
were unable to attract enough matter to build their own galaxies.\\

\noindent
3. New more precise nuclear chronology with several nuclear chronometers based on several long-lived isitopes,
revealed unexpectedly old stars in the Galaxy. 
Many of them have the age larger than the age of the Galaxy and one is even 
"older than the universe"~\cite{H-Bond} with the age
 ${14.46 \pm 0.31 }$ Gyr. The central age value exceeds the universe age by two standard deviations 
if  ${H= 67.3}$ km/sec/Mpc, as determined  from the angular fluctuations of the 
of the cosmic microwave background radiation, and hence ${t_U =13.8}$ Gyr, 
but if ${H= 74}$ km/sec/Mpc determined by the canonical  astronomical methods,
and so
${ t_U = 12.5}$ Gyr, the excess is by 9 standard deviations.  It is noteworthy that this stellar Methuselah
have large velocity, which may indicate its pregalactic origin. 

In the model of refs. \cite{ADJS,DKK} the long age found from the nuclear chronology may be mimicked  by the unusual 
initial chemical content of the star enriched by heavy elements.\\

\noindent
4. MACHOs  are low luminosity (in fact invisible) a half solar mass objects discovered by gravitational microlensing, for
a brief review see ref. \cite{BDK} . The  origin of these objects is mysterious. In principle they could be brown dwarfs or dead 
stars, but their density is too high to be created by the normal stellar evolution.  The only remaining possibility seems to be
primordial massive black holes. \\

\noindent
{ 5. The mass spectrum of the black holes in the Galaxy shows an unexpected maximum at 
${ (7.8 \pm 1.2) M_\odot }$.~\cite{M-BH-narrow}. This result agrees with other observations ~\cite{kreidberg}
demonstrating maximum at $M\sim{8M_\odot}$, with sharp decrease above ${10M_\odot}$ and practical absence
of BH  below ${5M_\odot}$.
Such behavior is unexplainable if BHs  are formed through  the stellar collapse, as it is usually assumed,  but nicely
fits the hypothesis of abundant  primordial BHs in the Galaxy.\\

\noindent
6. During last year several events of gravitational waves (GW) were registered at LIGO and Virgo interferometers, which
originated from the coalescence of massive  black hole binaries. 
The analysis of these event demonstrates beautiful agreement with General Relativity, in particular, for strong gravitational
fields near the horizon of the Schwarzschild black hole. However the origin and properties of the GW sources remains
mysterious. In majority of events the masses of the progenitor black holes are about $(20-30) M_\odot$, their spins  are
compatible with zero, and the spin of the baby BH created as a result of the coalescence is high as it is normally expected.
The following three problems of the standard theory emerged together with the GW discovery:\\
a) What is the origin of such heavy BHs, with ${\sim 30 M_\odot}$?\\
b) How the low spin values of the  progenitors can be explained?\\
c) How the binary system of BH can come from, presumably, stellar binary?\\
These problem are addressed in refs.~{ \cite{BDPP,talks-KAP,talk-AD}.
They are all  neatly solved if the initial black holes are primordial.\\

\noindent
{7. Intermediate mass BHs: ${M \sim 10^3 M_\odot}$, in globular clusters and ${M\sim 10^{4-5}}$ in
dwarf galaxies.}  Though the observation of the intermediate mass BH in globular clusters (GC) is not
an easy task,  there are signatures that indeed IMBHs with masses up to a few $10^4 M_\odot$ can be present in 
GC cores~\cite{gc-bh-1,gc-bh-2}.
The impact of such BH on the GC formation was studied in our paper~\cite{AD-KP}, where it was argued that 
PBH with masses about thousand solar mass may successfully describe GC formation in the observed amount.
We also mentioned that dwarf galaxies may be seeded by somewhat heavier PBH. A close point of view was also 
expressed in ref.~\cite{silk-dwarf}.

Recently an avalanche of black holes with masses in the range $(10^4 - 10^5) M_\odot$ appeared in the
sky, almost a thousand of them have been discovered~\cite{imbh-1,imbh-2,imbh-3}. It is difficult to find  any explanation
except for them being primordial. \\

\noindent
8. There are accumulating data indicating that SMBH in close vicinity to each other
are not so rare as it is naively expected: \\
a) Four binaries of SMBH are observed~\cite{bh-bin-1,bh-bin-2, bh-bin-3,bh-bin-4}.\\
b) Recently an observation of  four quasars in a narrow range in the sky
(quasar quartet) was reported~\cite{quartet}. According to the authors the probability for formation of such system 
by conventional mechanism is $10^{-7}$.
\\
c) In ref.~\cite{triple} a kiloparsec-scale {\it triple} supermassive black hole system at $z=0.256$ is
discovered.\\
The probability of an accidental formation of several supermassive black holes in close vicinity by the 
usual astrophysical processes is quite low, seemingly much lower than the accidental formation of close primordial
black holes.
\\

This list does not exhaust all mysteries in the sky. There are plenty of other data laeding to the same
conclusion of abundant population of the universe with massive primordial black hole. The new data in favor of
it appear practically every week and with new much more sensitive astronomical instruments the flux 
of new discoveries will increase multifold. It will allow ether to finally prove or disprove the discussed here picture.

\section{The mechanism of massive PBH creation}

{All these problems are uniquely and simply solved by creation in the early universe
of massive PBHs and compact stellar-like
objects. Such a mechanism was put forward a quarter of century ago~\cite{ADJS}.  In contrast to the previously considered 
mechanisms of PBH formation, the suggested mechanism leads to a rather wide-spread mass spectrum 
of PBHs. The spectrum is practically model independent and has the simple log-normal form:
\begin{eqnarray}
\frac{dN}{dM} = \mu^2 \exp{[-\gamma \ln^2 (M/M_0)], }
\label{dn-dM}
\end{eqnarray}
with only 3 constant parameters: ${\mu}$, ${\gamma}$,  ${M_0}$.

During recent years such mass distribution of PBH was rediscovered in several works and became quite popular.

The mechanism of ref.~\cite{ADJS} (see also development in ref.~\cite{DKK}) is based on a slightly modified 
Affleck-Dine (AD) \cite{Aff-Dine} scenario of baryogenesis. In the original version the scenario was based on  supersymmetry,
but the full supersymmetry (SUSY) is not necessary, we need only a few simple and natural features out of it.
SUSY predicts an existence of scalar fields, $\chi$ with nonzero baryonic number,  ${ B_\chi \neq 0}$. A generic feature of the 
SUSY models is an existence of the so called flat directions in the potential $U_\chi \equiv U(\chi, \chi^*)$, i.e. the directions
in the field space along which the potential does not rise. The field $\chi$
can condense along such flat directions and acquire quite a large value. It indeed happened during inflation due to 
rising quantum fluctuations of light scalars with mass smaller than the Hubble parameter, $H_I$. After the end of inflation
this large value of $\chi$ is transformed into large baryonic number of other particles, say,  quarks.

We consider the toy model with the potential:
\begin{eqnarray} 
U(\chi,\chi^*) =  \frac{1}{2} \left[ \lambda (\chi^4 + \chi^{*4}) + m^2 \chi^2 + m^{*\,2}\chi^{*\,2}\right] = \\ \nonumber
\lambda |\chi|^4 \left( 1+ \cos 4\theta \right) + |m|^2 |\chi|^2| \left[ 1+ \cos (2\theta+2\alpha)\right],
\label{U-of-chi}
\end{eqnarray}
where ${ \chi = |\chi| \exp (i\theta)}$ and ${ m=|m|e^\alpha}$. If ${\alpha \neq 0}$, C and CP are  broken.
In supersymmetric Grand Unified Theories (SUSY GUT) the  baryonic number of $\chi$ is naturally non-conserved. It is 
imposed  in expression~(\ref{U-of-chi}) as an absence of invariance with respect to the phase transformation of $\chi$.
 
{Initially (after inflation) the homogeneous field ${\chi}$ is away from origin and, when 
inflation is over, it starts to evolve down to the  equilibrium point, ${\chi =0}$,
according to the equation of the Newtonian mechanics:}
\begin{eqnarray} 
\ddot \chi +3H\dot \chi +U' (\chi) = 0.
\end{eqnarray}
The baryonic number of $\chi$ is the mechanical angular momentum in the two-dimensional complex plane of $[\chi,\chi^*]$:
\begin{eqnarray} 
B_\chi =\dot\theta |\chi|^2
\end{eqnarray}

The decay of ${{\chi}}$ transferred its baryonic number to that of quarks in B-conserving process. Due to possible initial  large 
amplitude of $\chi$\ the AD baryogenesis could lead to the cosmological  baryon asymmetry of order of unity, much larger than 
the observed value $\beta = n_B/n_\gamma \approx{6\times 10^{-10}}$.

If ${ m\neq 0}$, the angular momentum or, what is the sane, B, is generated because of different 
direction of the  quartic and quadratic valleys at low ${\chi}$, see more below.

In the papers~\cite{ADJS,DKK} the potential of $\chi$ was modified by an addition of the general renormalizable 
coupling to the inflaton field $\Phi$, the first term in the equation below:
\begin{eqnarray} 
U = {g|\chi|^2 (\Phi -\Phi_1)^2}  +
\lambda |\chi|^4 \,\ln \left[ \frac{|\chi|^2 }{\sigma^2 }\right]
{{+\lambda_1 \left(\chi^4 + h.c.\right) + %\nonumber\\
(m^2 \chi^2 + h.c.). \,\,\,\,\,\,\,\,\,\,\,\,\,\,\,\,\,\,\,\,
}}
\end{eqnarray}
It is assumed that the inflaton field $\Phi$ takes the value $\Phi_1$ sufficiently  long before the end of inflation
such that about 30-40 e-folding still remains.

The coupling to the inflaton ensures that the gates to the flat directions are open only for a short period of time during which
$\Phi$ is sufficiently close to $\Phi_1$. So we expect the following picture. If the window to flat direction, when ${\Phi \approx \Phi_1}$,
is open only for a sufficiently  short period, then with a small probability the bubbles with a high baryonic 
density could be formed. They would occupy a minor fraction of the universe volume, while the rest of the universe would 
have the normal small
value of the baryon asymmetry,  {${{ \beta \approx 6\cdot 10^{-10}}}$, created by small ${\chi}$}, which was not able to penetrate  
trought he gate to a large distance 
from the origin. These bubbles could be called High B Bubbles or HBB, it should not be confused with the nick name of the
 American coffee, Hot Brown Beverage. Though cosmologically small, HBBs can be astronomically large, having masses of the 
 order of the solar mass or even much higher. 
 As it is already mentioned, the universe would look as Swiss cheese, where the holes are the HBBs,
which may have much larger total mass than the mass of the rest of matter in the universe.
This process of the HBB formation can be called the phase transition of the 3/2 order.

At the high temperature cosmological epoch, when the electroweak  (EW) symmetry was unbroken, quarks were massless and a 
large contrast in the value of the baryonic number density between the HBB and the bulk of the universe did not lead to 
noticeable density perturbations.
These are the so called isocurvature perturbation. Even after the EW symmetry breaking the density contrast remained small and 
only after the QCD phase transition at the temperature $T = 100 - 200 $ MeV, when quarks formed heavy baryons, a high density
contrast between HBB and the bulk would appear. As a result, either primordial black holes or compact stellar-like objects 
would be created.

It is interesting that the sign of $\beta$ may be both negative and positive, It means that both baryonic and antibaryonic bubbles
can be formed. As a result PBH and anti-PBH could be created but they are indistinguishable (if there is no long-range forces 
associated with the baryon number). However compact stars and anti-stars  might be abundantly produced and could be in
a large amount even in the Galaxy. The relative density of these anti-objects is model dependent and can vary from zero 
to almost 50\%. 
Cosmologically large matter and antimatter domain may exist but not necessarily in equal amount, so  most probably the
global baryonic number of the universe is non-zero. 

The analysis of the observational bound on the compact objects in the Galaxy was performed in the 
papers~\cite{BDK,DB,BD}. It is a challenging problem to find anti-stars in the Galaxy.

\section{Conclusion \label{s-concl}}

The described mechanism~\cite{ADJS,DKK} nicely fits the basic trend of the surprising astronomical data
reviewed in Sec.~\ref{s-rev-obs}. In a sense this picture of the universe was predicted a quarter of century ago.
There are still quite many problems demanding deeper research:\\
$\bullet $ Quantitative study of the galaxy formation seeded by massive PBH.
In particular it would be interesting to find how the mass of the seed 
influences the emerging galaxy type.  Such an analysis worth doing for large galaxies, dwarfs, and globular clusters. \\
$\bullet $ Formation of galaxies in low density regions of  the universe. Such a study could explain existence of  
small galaxies with superheavy black holes inside, or even existence of superheavy BH in almost empty space.\\
$\bullet $ The problem of high velocity stars in galaxies. What is the source of their acceleration? Are they primordial stars or
are accelerated by the abundant population of  the PBH with intermediate masses, $(10^3-10^5) M_\odot$?\\
$\bullet $ A possible outcome of the described here mechanism of massive PBH creation is an accompanying creation of antimatter
objects, which may be abundant in the Galaxy.  Their discovery would prove the validity of the mechanism.

\section*{Acknowledgments}
This work was supported by the RNF Grant Grant N 16-12-10037.
I  thank Harald Fritzsch for the invitation to the Conference and 
also Kok Khoo Phua for the hospitality at NTU, Singapore.


\begin{thebibliography}{10}

\bibitem{ZN}
Y. B. Zel'dovich and I. D. Novikov, 
%The Hypothesis of Cores Retarded during Expansion and the Hot Cosmological Model, 
{\it Sov. Astron.}, {\bf 10} (1967), 602.

\bibitem{ADJS}
A.Dolgov and  J.Silk, "Baryon isocurvature fluctuations at 
small scales and baryonic dark matter" 
{\em Phys. Rev.}  {\bf D47} 4244 (1993).

\bibitem{DKK}
A.D. Dolgov, M. Kawasaki,  N.~Kevlishvili, 
"Inhomogeneous baryogenesis, cosmic antimatter, and dark matter",
{\it Nucl.Phys.,}  {\bf B807}, 229 (2009).

\bibitem{AD-KP}
A. Dolgov, K. Postnov,
"Globular Cluster Seeding by Primordial Black Hole Population",    
{\em JCAP} {\bf 1704}  no.04, 036 (2017).

\bibitem{monsters}
A.D. Dolgov, 
"Beasts in Lambda-CDM Zoo", 
{\em Phys. Atom. Nucl.} {\bf 80} (2017) no.5, 987. 

\bibitem{AD-UFN}
A.D. Dolgov,  "Massive and supermassive black holes in the contemporary and early Universe and problems in 
cosmology and astrophysics",
{\em Usp. Fiz. Nauk}, {\bf 188} (2018) no.2, 121; {\em Phys.Usp.} {\bf 61} (2018) no.2, 115.  

\bibitem{QSO-10}
Xue-BingWu et al,
"An ultraluminous quasar with a twelve billion solar mass black hole at redshift 6.30".
{\em Nature}, {\bf 518}, 512 (2015).

\bibitem{accr-rate}
M.A. Latif, M Volonteri, J.H. Wise, 
"Early growth of typical high redshift black holes seeded by direct collapse".
arXive 1801.07685.

\bibitem{gal-hi-L}
Tsai C.-W. et al, 
%P.R.M. Eisenhardt, J. Wu, {\it et al}, arXiv:1410.1751.
{"The most luminous galaxies discovered by WISE".} 
{\em Astrophys. J}., {\bf 805} (2015) Issue 2, article id. 90, 15 pp. 

\bibitem{gal-dens}
D. Waters, {\it et al}, 
"Monsters in the Dark"
{\em Mon. Not. Roy. Astron. Soc.} {\bf 461} (2016), L51

\bibitem{dust-ctlg}
M. Mancini, et al, 
"The dust mass in $ z>6 $ normal star forming galaxies".
arXiv:1505.01841.

\bibitem{empty-BH}
J.J. Condon, et al., 
"A Nearly Naked Supermassive Black Hole",
{\em Astrophys. J.}, {\bf 834} (2017) 184; arXiv:1606.04067.

\bibitem{H-Bond}
H.E. Bond, {\it et al},
"HD 140283: A Star in the Solar Neighborhood that Formed Shortly After the Big Bang".
%E. P. Nelan, D. A. VandenBerg, G. H. Schaefer,  D. Harmer, 
{\it Astrophys. J. Lett.}, {\bf 765} (2013), L12. %L12; arXiv:1302.3180,
%"HD 140283: A Star in the Solar Neighborhood that Formed Shortly After the Big Bang".

\bibitem{BDK}
S.I. Blinnikov, A.D. Dolgov, K.A. Postnov,
"Antimatter and antistars in the universe and in the Galaxy",  
{\em Phys. Rev.}, {\bf D92} (2015), 023516.

\bibitem{M-BH-narrow}
F. Ozel  F., {\it et al.},
% D. Psaltis, R. Narayan, J.E. McClintock, 
"The Black Hole Mass Distribution in the Galaxy",
{\em Astrophys.J.} {\bf 725} (2010) 1918-1927; 
arXiv:1006.2834.

\bibitem{kreidberg}
L. Kreidberg, {\it et al},
%C.D. Bailyn, W.M. Farr, V. Kalogera,
"Mass Measurements of Black Holes in X-Ray Transients: Is There a Mass Gap?"
arXiv:1205.1805. %submitted to Ap.J.

\bibitem{BDPP}
D. Blinnikov, S. Dolgov, N. Porayko, K. Postnov, 
"Solving puzzles of GW150914 by primordial black holes", 
 {\em JCAP},  {\bf 1611} (2016), 036;  arXiv:1611.00541. 
 
 \bibitem{talks-KAP}
K.A. Postnov, lecture at International school of young astronomers
“MAGNETO PLASMIC PROCESSES IN RELATIVISTIC  ASTROPHYSICS”
Tarusa (Kaluzhsky region, Russia) 12-16 September, 2016.

\bibitem{talk-AD}
%A.D. Dolgov, Invited talk at the
%International Symposium Advances in Dark Matter and Particle Physics (ADMPP16), 24-27 Oct 2016. Messina, Italy,
%arXiv:1701.05774;\\
%\bi{ad-talks}
A.D. Dolgov,
"Problems with the sources of the observed gravitational waves and their resolution",
 Invited talk at the
International Symposium Advances in Dark Matter and Particle Physics (ADMPP16), 24-27 Oct 2016. Messina, Italy,
 {\it EPJ Web Conf.}, {\bf 142} (2017), 01012; arXiv:1701.05774 [astro-ph.CO].

\bibitem{gc-bh-1}
B. Kızıltan, H. Baumgardt,  A. Loeb, 
"An intermediate-mass black hole in the centre of the globular cluster 47 Tucanae", 
{\it Nature}, {\bf 542} (Feb., 2017) 203–205.

\bibitem{gc-bh-2}
H. Baumgardt, 
"N -body modelling of globular clusters: masses, mass-to-light ratios and intermediate-mass black holes", 
{\it MNRAS},  {\bf 464} (Jan., 2017) 2174–2202,

\bibitem{silk-dwarf}
J. Silk, 
"Feedback by Massive Black Holes in Dwarf Galaxies",
arXiv:1703.08553

\bibitem{imbh-1}
M. Mezcua, {\it et al}
"Intermediate-mass black holes in dwarf galaxies out to redshift  $\sim 2.4$ in the Chandra COSMOS Legacy Survey"
arXiv:1802.01567 [astro-ph.GA].

\bibitem{imbh-2}
I.V. Chilingarian, 
"A Population of Bona Fide Intermediate Mass Black Holes Identified as Low Luminosity Active Galactic Nuclei",
arXiv:1805.01467.

\bibitem{imbh-3}
He-Yang Liu, et al, A Uniformly Selected
"Sample of Low-Mass Black Holes in Seyfert 1
Galaxies. II. The SDSS DR7 Sample",
arXiv:1803.04330,

\bibitem{neutral}
E. Ba{$\tilde n$}ados {\it et al}
 "An 800-million-solar-mass black hole in a significantly neutral Universe at redshift 7.5",
arXive: 1712.01860.

\bibitem{super-accr}
Yongjung Kim, {\it et al}
"The Infrared Medium-deep Survey. IV. Low Eddington Ratio of A Faint Quasar at z∼6: Not Every Supermassive 
Black Hole is Growing Fast in the Early Universe",
arXive 1802.02782.

\bibitem{bh-bin-1}
P. Kharb, et al "A candidate sub-parsec
binary black hole in the Seyfert galaxy
NGC 7674", $ d=116$ Mpc, %$M= 3:63\times 10^7 M_\odot$
arXive 1709.06258.

\bibitem{bh-bin-2}
C. Rodriguez {\it et al}. 
"A compact supermassive binary black hole system." 
{\it Ap. J.} {\bf 646}, 49 (2006), %$ d =230 $ Mpc

\bibitem{bh-bin-3}
M.J. Valtonen, 
"New orbit solutions for the precessing binary black hole model of OJ 287", 
{\it Ap.J.}, {\bf 659}, 1074 (2007). %$z \approx 0.3$

\bibitem{bh-bin-4}
M.J. Graham {\it et al}. 
"A possible close supermassive black-hole binary in a quasar with optical periodicity". 
{\it Nature},  {\bf 518}, 74 (2015). %$ z \approx  0.3$.

\bibitem{quartet}
Hennawi  J.F.,  { et al}, {\it Science}, {\bf 348} (2015), 779.

\bibitem{triple}
E.i Kalfountzou, M. Santos L.M. Trichas
"SDSS J1056+5516: A Triple AGN or an SMBH Recoil Candidate?"
arXiv:1712.03909.

\bibitem{Aff-Dine}
 I. Affleck, M. Dine, 
"A New Mechanism for Baryogenesis", 
 {\it Nucl. Phys}. {\bf B 249}, 361 (1985).


\bibitem{DB}
A. D. Dolgov, S. I. Blinnikov, 
"Stars and Black Holes from the very Early Universe",
{\it Phys. Rev.} {\bf D. 89}, 021301 (2014), arXiv:1309.3395

\bibitem{BD}
C. Bambi, A. D. Dolgov, {\it Nuclear Physics}, 
"Antimatter in the Milky Way", 
{\bf B 784}, 132 (2007).

\end{thebibliography}
\end{document}